   \documentstyle[psfig]{l-aa}
\def \etal     {et al.}
\def \ie       {i.\,e.}
\def \eg       {e.\,g.}
\def \vLSR     {\hbox{${v_{\rm LSR}}$}}
\def \TMB      {\hbox{$T_{\rm MB}$}}
\def \Tkin     {\hbox{$T_{\rm kin}$}}
\def \kms      {\hbox{${\rm km\,s}^{-1}$}}                % km/s
\def \Kkms     {\hbox{${\rm K\,km\,s}^{-1}$}}             % K*km/s
\def \arcdeg   {\hbox{$^{\circ}$}}                        % d
\def \arcmin   {\hbox{$^\prime$}}                         % '
\def \arcsec   {\hbox{$^{\prime\prime}$}}                 % "
\def \RA#1     {\hbox{$\alpha_{#1}$}}                     % R.A.
\def \Dec#1    {\hbox{$\delta_{#1}$}}                     % decl.
\def \twCO     {\hbox{$^{12}$CO}}                         % 12CO
\def \thCO     {\hbox{$^{13}$CO}}                         % 13CO
\def \CeiO     {\hbox{C$^{18}$O}}                         % C18O
\def \CthS     {\hbox{C$^{33}$S}}                         % C33S
\def \CfoS     {\hbox{C$^{34}$S}}                         % C34S
\def \twCN     {\hbox{$^{12}$CN}}                         % 12CN
\def \thCN     {\hbox{$^{13}$CN}}                         % 13CN
\def \FORM     {\hbox{H$_2$CO}}                           % H2CO
\def \HtwCN    {\hbox{H$^{12}$CN}}                        % H12CN
\def \HthCN    {\hbox{H$^{13}$CN}}                        % H13CN
\def \HCfoN    {\hbox{HC$^{14}$N}}                        % HC14N
\def \HCfiN    {\hbox{HC$^{15}$N}}                        % HC15N
\def \HNthC    {\hbox{HN$^{13}$C}}                        % HN13C
\def \HCOp     {\hbox{HCO$^+$}}                           % HCO+
\def \HtwCOp   {\hbox{H$^{12}$CO$^+$}}                    % H12CO+
\def \HthCOp   {\hbox{H$^{13}$CO$^+$}}                    % H13CO+
\def \Cp       {\hbox{C$^+$}}                             % C+
\def \C#1      {\hbox{$^{#1}$C}}                          % C isotopes
\def \ISOC     {\hbox{$^{12}$C$/^{13}$C}}                 % 12C/13C      \ISOC
\def \ISON     {\hbox{$^{14}$N$/^{15}$N}}                 % 14N/15N      \ISON
\def \ISOA     {\hbox{$^{16}$O$/^{18}$O}}                 % 16O/18O      \ISOA
\newcommand{\see}[1]{$^{\rm #1)}$}
\def \JMolSp   {{\rm J. Mol. Spec.}}
\def \bra#1    {$\left\{\makebox{\rule[-#1ex]{0pt}{#1ex}}\right.$}
\def \ket#1    {$\left.\makebox{\rule[-#1ex]{0pt}{#1ex}}\right\}$}
\def \uspace#1 {\makebox{\rule[#1ex]{0pt}{2ex}}}
\def \dspace   {\makebox{\rule[-2ex]{0pt}{2ex}}}

\begin{document}

\thesaurus{3(09.01.1, 11.01.1, 13.19.1, 13.19.3)}

\title{Dense gas in nearby galaxies}

\subtitle{XI. Interstellar $^{12}$C/$^{13}$C ratios in the central regions of
          M\,82 and IC\,342{\thanks{Based on observations with the IRAM 30-m
                   telescope, Pico Veleta, Spain and the Swedish-ESO
                   Submillimetre Telescope (SEST) at the European Southern
                   Observatory (ESO), La Silla, Chile} } }
\author{
   C.~Henkel\inst{1}, Y.-N.~Chin\inst{2,3}, R.~Mauersberger\inst{4},
   \and J.B.~Whiteoak\inst{5,6}
}

\offprints{C.~Henkel, p220hen@mpifr-bonn.mpg.de}

\institute{
   Max-Planck-Institut f\"ur Radioastronomie,
   Auf dem H\"ugel 69, D-53121 Bonn, Germany
\and
   Institute of Astronomy and Astrophysics, Academia Sinica,
   P.O.Box 1-87 Nankang, 115 Taipei, Taiwan
\and
   Radioastronomisches Institut der Universit\"at Bonn,
   Auf dem H\"ugel 71, D-53121 Bonn, Germany
\and
   Steward Observatory, The University of Arizona,
   Tucson, AZ\,85721, U.S.A.
\and
   Australia Telescope National Facility, Radiophysics Laboratories,
   P.O. Box 76, Epping, NSW 2121, Australia
\and
   Paul Wild Observatory, Australia Telescope National Facility, CSIRO,
   Locked Bag 194, Narrabri NSW 2390, Australia
}

\date{Received 9 September 1996 / Accepted 21 July 1997}
\maketitle

\begin{abstract}

   $^{12}$C$/^{13}$C line intensity ratios have been derived from several
   carbon-bearing molecules to confine the range of carbon and oxygen isotope
   abundance ratios toward the nuclear regions of two infrared bright galaxies.
   The most stringent limits are obtained from CN mm-wave emission lines.
   Supplementary measurements toward the Galactic center region indicate
   that overall $I$(\twCN)/$I$(\thCN) line intensity ratios are giving
   lower limits to the corresponding \ISOC\ abundance ratio.
   Toward M\,82 and IC\,342, we find \ISOC\ $>$ 40 and $>$30.
   Therefore the smaller \ISOC\ ratio of our own Galactic center region (25)
   may not be typical for central regions of galaxies which are more
   luminous in the infrared.
   The \ISOC\ limits and data from various isotopic species of CO also infer
   \ISOA\ abundance ratios of $>$ 90 and $>$ 125 for M\,82 and IC\,342,
   respectively.
   From the $I$(\HCfoN)/$I$(\HCfiN) line intensity ratio, \ISON\ $>$ 100
   is derived for M\,82.

\keywords{ISM: abundances -- Galaxies: abundances -- Radio lines: ISM --
          Radio lines: galaxies}

\end{abstract}

\section{Introduction}
\label{sec:CN-Introduction}

   Isotope ratios derived from CNO elements have significantly contributed
   to our understanding of the nuclear processing in stars and the `chemical'
   evolution of galaxies, since these elements are abundant and have stable
   `primary' and `secondary' nuclei.
   From a theoretical point of view, \ISOC\ is the
   least controversial CNO isotope ratio:
   \C12 \ is a `primary' product of helium-burning, \C13 \ is mainly
   a `secondary' product of hydrogen-burning with \C12 \ as the seed nucleus.
   Some primary \C13 \ may also be synthesized during the third dredge up in
   stars of intermediate mass (`hot bottom burning', \eg\ Renzini \& Voli 1981).

   There is evidence for high \ISOC\ ratios in the central regions of
   active star-forming galaxies with high luminosities in the far infrared
   (for a summary, see Henkel \& Mauersberger 1993).
   First hints were obtained from distant mergers (being ultraluminous in
   the infrared) which were showing integrated $I$(\twCO)/$I$(\thCO) $J$=1--0
   line intensity ratios $>$20 (Aalto \etal\ 1991;
   Combes \etal\ 1991; Casoli \etal\ 1992a,b).
   An interpretation in terms of an extended halo of weak $^{12}$CO
   and negligible $^{13}$CO emission is not supported by a comparison of
   filled-aperture with interferometric CO data (P.M. Solomon, priv. comm.).
   The generally accepted explanation involves inflow of disk gas
   with high \ISOC\ ratios into the central region, possibly combined with
   a \C12 \ enhancement caused by nucleosynthesis in massive stars.
   Most direct evidence for high \ISOC\ ratios was obtained from
   recent studies of the central regions of the nearby starburst galaxies
   NGC\,253 and NGC\,4945; this is based on \ISOC\ line intensity ratios
   from a variety of molecular species (Henkel \etal\ 1993, 1994).
   The estimated \ISOC\ abundance ratios, 40 -- 50, are larger than
   the value for the central region of the Milky Way, $\sim$ 25
   (\eg\ Wilson \& Rood 1994).

   Since two well studied extragalactic sources represent too small
   a sample for a comparison with the Milky Way, we have extended this
   list, including M\,82 (NGC\,3034) and IC\,342.
   These contain powerful far infrared sources in their central regions
   and show an impressive amount of strong molecular lines
   (\eg\ Henkel \etal\ 1986, 1991).
   From our experience with NGC\,253, the best limits to the \ISOC\ abundance
   ratio are obtained from the $I$(\twCN)/$I$(\thCN) line intensity ratios.
   We observed CN not only toward M\,82 and IC\,342
   but also toward the Galactic center region (see Table \ref{tbl:CN-source}),
   where the interstellar \ISOC\ ratio is known and where isotope ratios
   deduced from CN can thus be tested.

\section{Observations}
\label{sec:CN-Observations}

\subsection{30-m IRAM telescope}
\label{sec:CN-iram}

   Observations of CN, HCN, \HCOp, and HNC (for frequencies, see Tables
   \ref{tbl:CN-line} and \ref{tbl:CN-ISO}) were made toward the central
   regions of M\,82 and IC\,342 in June 1993 using the IRAM 30-m telescope
   (Baars \etal\ 1987) on Pico Veleta, Spain.
   Between 86 and 113\,GHz the full width to
   half power beam size was 26\arcsec\ -- 22\arcsec.
   An SIS receiver was employed with system temperatures of $\sim$ 500\,K
   on a main beam brightness temperature (\TMB; \eg\ Downes 1989) scale.
   The $\lambda$ = 3\,mm main beam efficiency was $\eta_{\rm MB}$ $\sim$ 0.60
   As backend an 864 channel acousto-optical spectrometer (AOS) with a
   channel spacing of 0.58\,MHz was used for \twCN\@.
   A 512$\times$1\,MHz filterbank was used for \thCN, \HtwCOp, \HthCOp,
   \HtwCN, \HthCN, \HCfiN, and \HNthC, respectively.
   The `channel spacing' ($\sim$ 90\% of the `channel resolution')
   corresponds to a velocity separation of 1.5\,\kms\ for
   the \twCN\ $N$=1--0 transition and 2.8 -- 3.5\,\kms\ for the other lines.
   The spectra were obtained with a wobbling secondary mirror using
   a switch cycle of four seconds (2\,s on source, 2\,s off-source)
   and a beam throw of 200\arcsec\ -- 240\arcsec\ in azimuth.
   The pointing accuracy, based on nearby continuum sources and
   the lineshape of the simultaneously measured \twCO(2--1) feature,
   was $\sim$ 5\arcsec or better.

   All data were calibrated by observing a cold load at liquid nitrogen
   temperature and a chopper wheel at ambient temperature.
   The receivers were tuned to a single sideband mode with an
   image sideband rejection of typically 7\,dB.
   The image sideband rejection was determined with a signal generator,
   inducing lines at the signal and image sideband frequencies about
   one or two days prior to the start of the observations.
   During the observations line intensities were regularly
   monitored toward Orion-KL and IRC+10216.
   These remained stable to within 10 -- 15\% of the values
   given by Mauersberger \etal\ (1989).

\subsection{15-m SEST}
\label{sec:CN-sest}

   The measurements toward the Galactic center region were
   made in May 1994 using the 15-m Swedish-ESO Submillimetre Telescope
   (SEST; Booth \etal\ 1989) at La Silla, Chile.
   At the frequencies of the $N$=1--0 transitions of
   \twCN\ and \thCN\ the beamwidth was $\sim$ 45\arcsec.
   The main-beam efficiency was $\eta_{\rm MB}$ = 0.70.
   We employed a Schottky receiver (which was sensitive to only one sideband)
   with system temperatures of order 500\,K on a \TMB\ temperature scale.
   The image sideband rejection was $\sim$ 15\,dB.
   The backend was an acousto-optical spectrometer with 1600 contiguous
   channels and a channel separation of 0.68\,MHz (1.8 -- 1.9\,\kms).
   All measurements were carried out in position-switching mode.
   The reference position ($l^{\rm II}$,$b^{\rm II}$) = (0\arcsec,+720\arcsec)
   was checked to have no emission down to a 3$\sigma$ level of \TMB\ = 50\,mK.
   No absorption, caused by emission from the off-positions,
   was seen in any of the Galactic center spectra.
   Integration times varied from 1 minute for \twCN\ to 20 minutes for \thCN.

   The pointing accuracy, obtained from measurements of SiO maser sources,
   was better than 10\arcsec.
   The stability of the 3\,mm Schottky receiver was examined by comparing
   \CfoS\ and \CthS\ spectra from Orion-KL, taken in September 1993, May
   1994, and January 1995; the line temperatures are consistent within 15\%.

\begin{table}
   \caption[]
           {Source list}
 \label{tbl:CN-source}
   \begin{flushleft}
   \begin{tabular}{l r@{\,}l@{~~}r@{~~}r@{~}c}
   \hline
   \uspace1
   Object        & \multicolumn{2}{c}{Distance}  & \multicolumn{1}{c}{\RA1950 }
                 & \multicolumn{1}{c}{\Dec1950 } &  \vLSR \\
   \dspace       & \multicolumn{2}{c}{} & \multicolumn{1}{c}{[$^{\rm h\ m\ s}$]}
                 & \multicolumn{1}{c}{[\arcdeg\ \arcmin\ \arcsec]}  & [\kms] \\
   \hline
   \uspace1
   IC\,342       & ~1.8 & Mpc  &   3~41~57.5  &    67~56~40  &  ~40  \\
   M\,82         & ~3.3 & Mpc  &   9~51~43.0  &    69~55~00  &  180  \\
   \uspace0.5
   M$-$013$-$008 & ~8.5 & kpc  &  17~42~26.5  & $-$29~04~10  &  ~15  \\
   M$-$002$-$007 & ~8.5 & kpc  &  17~42~40.0  & $-$28~58~14  &  ~50  \\
%  M$+$010$-$001 & ~8.5 & kpc  &  17~42~43.3  & $-$28~50~13  &  ~50  \\
   M$+$007$-$008 & ~8.5 & kpc  &  17~42~55.3  & $-$28~53~57  &  ~50  \\
   \dspace
   M$+$011$-$008 & ~8.5 & kpc  &  17~43~01.1  & $-$28~51~55  &  ~50  \\
   \hline
   \end{tabular}
   \end{flushleft}
\ \ \\
\end{table}

\begin{figure}
   \vspace*{-12 mm}
   \hspace*{-6 mm} \psfig{figure=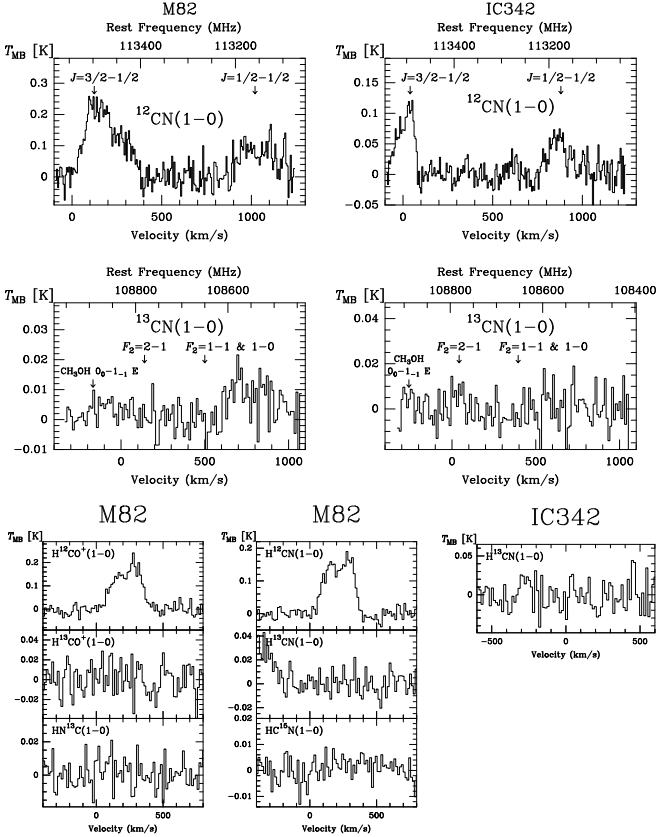,width=10 cm}
   \vspace*{-10 mm}
   \caption[]
           {Observed spectra of molecules and their \C13 -bearing isotopic
            species toward M\,82 and IC\,342.
            The spectra are smoothed to a channel spacing of 4\,MHz
            ($\sim$ 12\,\kms).}
 \label{fig:CN-Galaxies}
\end{figure}

\section{Results}
\label{sec:CN-Results}

   Distances, coordinates, and radial velocities of the observed sources are
   displayed in Table \ref{tbl:CN-source}.
   All data were converted to a $T_{\rm MB}$ scale and
   first order baselines were subtracted.
   The CN line parameters can be found in Table \ref{tbl:CN-line}.
   Fig.\,\ref{fig:CN-Galaxies} displays the spectra obtained toward
   the nuclear regions of M\,82 and IC\,342.
   Toward these extragalactic sources, the \twCN\ hyperfine components
   (see Sect.\,\ref{sec:CN-hyperfine}) are not resolved and only two broad
   features are seen.
   \thCN\ was not detected.
   The broad feature in the \thCN\ spectrum of M\,82 (rest frequency
   either 108.58 (signal sideband) or 111.82\,GHz (image sideband) for
   \vLSR\ $\sim$ 180\,\kms ) remains unidentified.
   While IC\,342 has also been observed in HCN $J$=1--0, M\,82 has been
   observed additionally in the $J$=1--0 transitions of \HCOp, HCN, and HNC.
   A summary of line intensities, including a first \HCfiN\ spectrum from M82,
   can be found in Table \ref{tbl:CN-ISO}.

\begin{figure}
   \vspace*{-22 mm}
   \hspace*{-17 mm} \psfig{figure=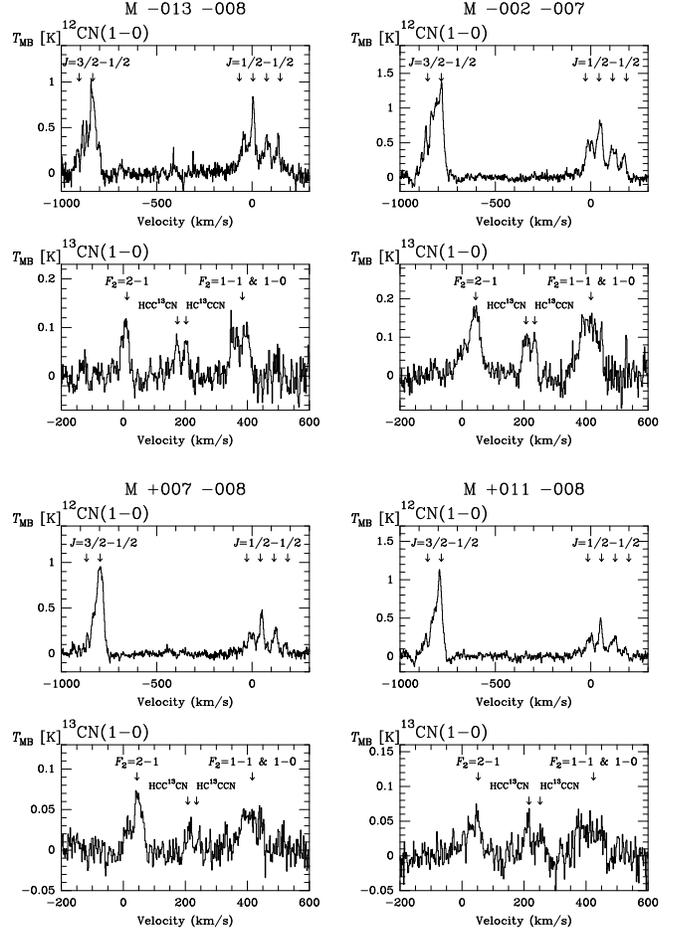,width=12 cm}
   \vspace*{-15 mm}
   \caption[]
           {Observed \twCN\ and \thCN\ spectra toward the Galactic center.}
 \label{fig:CN-GC}
\end{figure}

   $^{12}$CN($N$=1--0) and \thCN($N$=1--0) spectra toward sources in the
   Galactic center region are shown in Fig.\,\ref{fig:CN-GC}.
   The hyperfine structure of the \twCN\ $N$=1--0 $J$=3/2--1/2 spin-doublet
   line cannot be resolved, but it is no problem to resolve the hyperfine
   components of the $N$=1--0 $J$=1/2--1/2 transition (for details on
   fine structure and hyperfine splitting, see Sect.\,\ref{sec:CN-hyperfine}).
   For \thCN, the low resolution AOS can resolve the $N$=1--0
   hyperfine components from the Galactic center region into two groups.
   While one contains all the $F_1$=1 $F_2$=2--1 transitions, the other
   includes the $F_1$=1 $F_2$=1--1 and $F_1$=0 $F_2$=1--0 transitions.
   The $F_1$=1 $F_2$=0--1 transitions are weaker and remain undetected.

\begin{table*}
   \caption[]
           {\twCN\ to \thCN\ ratio toward sample in the Galactic center,
            M\,82, and IC\,342.}
 \label{tbl:CN-line}
   \begin{flushleft}
   \begin{tabular}{l l@{~}l@{~}c@{}l r c@{\hspace*{-2mm}}c@{~}r@{}l@{~$\pm$~}l
                   r@{}l@{~$\pm$~}l}
   \hline
   \uspace2
   Object  & \multicolumn{4}{l}{Molecule \& Transition}
           & \multicolumn{1}{c}{Frequency}
           & \multicolumn{5}{c}{$\int$ \TMB\,d$v$}
           & \multicolumn{3}{c}{$\frac{\hbox{$I$}(\twCN)}{\hbox{$I$}(\thCN)}$}\\
   \dspace & \multicolumn{4}{l}{}
           & \multicolumn{1}{c}{[GHz]}
           & \multicolumn{5}{c}{[\Kkms]}
           & \multicolumn{3}{c}{} \\
   \hline
   \uspace3
   M\,82   & \twCN & $N$=1--0 & \bra1 & \begin{tabular}{l} $J$=3/2--1/2 \\
                     $J$=1/2--1/2 \end{tabular}
           & \begin{tabular}{r} $\sim$ 113.491 \\ $\sim$ 113.180 \end{tabular}
           & \begin{tabular}{r@{.}l@{$\pm$}l} 52&3   & 5.8  \\ 22&4~  & 3.6~
             \end{tabular} & \ket1          & 74&.7  & 6.9
           &  \multicolumn{3}{c}{$>$ 42} \\
           & \thCN & $N$=1--0 &       &
           & \begin{tabular}{r} $\sim$ 108.720 \end{tabular}
           & & & $<$ 1&\multicolumn{2}{l}{\hspace*{-2mm}.8 \see{a}} \\
   \uspace3
   IC\,342 & \twCN & $N$=1--0 & \bra1 & \begin{tabular}{l} $J$=3/2--1/2 \\
                     $J$=1/2--1/2 \end{tabular}
           & \begin{tabular}{r} $\sim$ 113.491 \\ $\sim$ 113.180 \end{tabular}
           & \begin{tabular}{r@{.}l@{$\pm$}l} 11&2   & 2.3  \\  7&71  & 0.73~
             \end{tabular} & \ket1         &  18&.9  & 2.4
           &  \multicolumn{3}{c}{$>$ 31} \\
           & \thCN & $N$=1--0 &       &
           & \begin{tabular}{r} $\sim$ 108.720 \end{tabular}
           & & & $<$ 0&\multicolumn{2}{l}{\hspace*{-2mm}.61 \see{b}} \\
   \uspace8
   M-013-008 & \twCN & $N$=1--0 & \bra6 & \begin{tabular}{l}
                       $J$=3/2--1/2 \see{c}      \\ $J$=1/2--1/2 $F$=3/2--3/2 \\
                       $J$=1/2--1/2 $F$=3/2--1/2 \\ $J$=1/2--1/2 $F$=1/2--3/2 \\
                       $J$=1/2--1/2 $F$=1/2--1/2 \end{tabular}
             & \begin{tabular}{r} $\sim$ 113.491 \\        113.191 \\
                       113.171 \\        113.144 \\        113.123 \end{tabular}
             & \begin{tabular}{r@{.}l@{$\pm$}l}     56&5   & 1.1  \\
                14&4   & 0.7  \\  23&6   & 0.7  \\  16&6   & 0.7  \\
                12&3~  & 0.8~~ \end{tabular}
                           & \ket6         & 123&    & 1.8  &  15&.0  & 0.9 \\
             & \thCN & $N$=1--0 & \bra1 & \begin{tabular}{l} $F_2$=2--1 \\
                       $F_2$=1--1 \& $F_2$=1--0 \end{tabular}
             & \begin{tabular}{r} $\sim$ 108.720 \\ $\sim$ 108.654 \end{tabular}
             & \begin{tabular}{r@{.}l@{$\pm$}l}      2&75  & 0.22 \\
                 5&47  & 0.29 \end{tabular}
                     & \ket1  &    8&.2  & 0.4   & \multicolumn{3}{c}{} \\
   \uspace8
   M-002-007 & \twCN & $N$=1--0 & \bra6 & \begin{tabular}{l}
                       $J$=3/2--1/2 \see{c}      \\ $J$=1/2--1/2 $F$=3/2--3/2 \\
                       $J$=1/2--1/2 $F$=3/2--1/2 \\ $J$=1/2--1/2 $F$=1/2--3/2 \\
                       $J$=1/2--1/2 $F$=1/2--1/2 \end{tabular}
             & \begin{tabular}{r} $\sim$ 113.491 \\        113.191 \\
                       113.171 \\        113.144 \\        113.123 \end{tabular}
             & \begin{tabular}{r@{}l@{$\pm$}l}     104&    & 0.7  \\
                26&.1  & 0.5  \\  30&.8  & 0.4  \\  17&.9  & 0.4  \\
                 9&.66 & 0.39~~ \end{tabular}
                           & \ket6         & 189&    & 1.1  &   9&.4  & 0.3 \\
             & \thCN & $N$=1--0 & \bra1 & \begin{tabular}{l} $F_2$=2--1 \\
                       $F_2$=1--1 \& $F_2$=1--0 \end{tabular}
             & \begin{tabular}{r} $\sim$ 108.720 \\ $\sim$ 108.654 \end{tabular}
             & \begin{tabular}{r@{.}l@{$\pm$}l}      9&34  & 0.38 \\
                10&7   & 0.4~~~ \end{tabular}
                     & \ket1  &   20&.0  & 0.6   & \multicolumn{3}{c}{} \\
   \uspace8
   M+007-008 & \twCN & $N$=1--0 & \bra6 & \begin{tabular}{l}
                       $J$=3/2--1/2 \see{c}      \\ $J$=1/2--1/2 $F$=3/2--3/2 \\
                       $J$=1/2--1/2 $F$=3/2--1/2 \\ $J$=1/2--1/2 $F$=1/2--3/2 \\
                       $J$=1/2--1/2 $F$=1/2--1/2 \end{tabular}
             & \begin{tabular}{r} $\sim$ 113.491 \\        113.191 \\
                       113.171 \\        113.144 \\        113.123 \end{tabular}
             & \begin{tabular}{r@{.}l@{$\pm$}l}     53&1   & 0.5  \\
                10&1   & 0.4  \\  13&6   & 0.3  \\   8&15  & 0.35 \\
                 2&06  & 0.27~ \end{tabular}
                           & \ket6         &  87&.0  & 0.8  &  15&.0  & 0.8 \\
             & \thCN & $N$=1--0 & \bra1 & \begin{tabular}{l} $F_2$=2--1 \\
                       $F_2$=1--1 \& $F_2$=1--0 \end{tabular}
             & \begin{tabular}{r} $\sim$ 108.720 \\ $\sim$ 108.654 \end{tabular}
             & \begin{tabular}{r@{.}l@{$\pm$}l}      2&65  & 0.16 \\
                 3&14 & 0.19 \end{tabular}
                     & \ket1 &     5&.8  & 0.3  & \multicolumn{3}{c}{}\\
   \uspace8
   M+011-008 & \twCN & $N$=1--0 & \bra6 & \begin{tabular}{l}
                       $J$=3/2--1/2 \see{c}      \\ $J$=1/2--1/2 $F$=3/2--3/2 \\
                       $J$=1/2--1/2 $F$=3/2--1/2 \\ $J$=1/2--1/2 $F$=1/2--3/2 \\
                       $J$=1/2--1/2 $F$=1/2--1/2 \end{tabular}
             & \begin{tabular}{r} $\sim$ 113.491 \\        113.191 \\
                       113.171 \\        113.144 \\        113.123 \end{tabular}
             & \begin{tabular}{r@{.}l@{$\pm$}l}     44&5   & 0.6  \\
                11&3   & 0.4  \\  13&9   & 0.4  \\  10&9   & 0.4  \\
                 2&20  & 0.32~ \end{tabular}
                           & \ket6         &  82&.7  & 1.0  &  13&.1  & 0.8 \\
   \uspace-3 & \thCN & $N$=1--0 & \bra1 & \begin{tabular}{l} $F_2$=2--1 \\
                       $F_2$=1--1 \& $F_2$=1--0 \end{tabular}
             & \begin{tabular}{r} $\sim$ 108.720 \\ $\sim$ 108.654 \end{tabular}
             & \begin{tabular}{r@{.}l@{$\pm$}l}      2&75  & 0.19 \\
                 3&56  & 0.24 \end{tabular}
                     & \ket1  &   6&.3  & 0.3  & \multicolumn{3}{c}{}\\
   \hline
   \end{tabular}
   \end{flushleft}
   {\footnotesize \begin{enumerate} \renewcommand{\labelenumi}{\alph{enumi})}
                                                  \renewcommand{\itemsep}{0pt}
   \item For undetected molecular lines, 3$\sigma$ upper limits
         of \hbox{$\int$\TMB\,d$v$} are given that are derived from the
         rms noise level.
   \item If the tentative feature observed is not representing \thCN\ emission,
         a 3\,r.m.s. limit analysis corresponding to that for M\,82
         yields $I$(\twCN)/$I$(\thCN) $>$ 31.
   \item The hyperfine components belonging to the \twCN\ $N$=1--0 $J$=3/2--1/2
         transition are not resolved in frequency.
   \end{enumerate} }
\end{table*}

\begin{table*}
   \caption[]
           {Isotopic ratios from molecular observations
            toward M\,82 and IC\,342}
 \label{tbl:CN-ISO}
   \begin{flushleft}
   \begin{tabular}{l l c c r@{$\pm$}l@{}c@{}c}
   \hline
   \uspace2 \uspace-3
   Galaxy  & Transition
           & \begin{tabular}{c} $\nu_{\rm ^{12}CX}$ \\ ~[MHz]   \end{tabular}
           & \begin{tabular}{c} $\nu_{\rm ^{13}CX}$ \\ ~[MHz]   \end{tabular}
           & \multicolumn{2}{l}{\begin{tabular}{c}
                                     $I$($^{12}$CX) \\ ~[\Kkms] \end{tabular}}
           & \begin{tabular}{c}      $I$($^{13}$CX) \\ ~[\Kkms] \end{tabular}
           & $\frac{\hbox{$I$($^{12}$CX)}}{\hbox{$I$($^{13}$CX)}}$ \\
   \hline
   \uspace1
   M\,82   & CN(1--0)    & \ldots \see{a} & \ldots \see{b} &~~74.4 & 6.9
                                          & $<$ 1.8 \see{e}&  $>$ 42 \see{e} \\
   \uspace0.5
           & \HCOp(1--0) &  89188.523     & 86754.330      &  44.6 & 1.3
                                          & $<$ 2.4 \see{e}&  $>$ 19 \see{e} \\
   \uspace0.5
           & HCN(1--0)   &  88631.602     & 86340.184      &  39.6 & 1.0
                                          & $<$ 1.7 \see{e}&  $>$ 23 \see{e} \\
   \uspace0.5
           & HNC(1--0)   &  90663.543     & 87090.859      &  13.4 & 0.8 \see{c}
                                          & $<$ 1.6 \see{e}&  $>$ ~9 \see{e} \\
   \uspace1
   IC\,342 & CN(1--0)    & \ldots \see{a} & \ldots \see{b} &  18.9 & 2.4
                                          & $<$ 0.6 \see{e}&  $>$ 31 \see{e} \\
   \uspace0.5 \dspace
           & HCN(1--0)   &  88631.602     & 86340.184      &  19.1 & 1.1 \see{d}
                                          & $<$ 2.4 \see{e}&  $>$ ~8 \see{e} \\
   \hline
   \uspace2 \uspace-3
         & & \begin{tabular}{c} $\nu_{\rm HC^{14}N}$ \\ ~[MHz]   \end{tabular}
           & \begin{tabular}{c} $\nu_{\rm HC^{15}N}$ \\ ~[MHz]   \end{tabular}
           & \multicolumn{2}{l}{\begin{tabular}{c}
                                     $I$(HC$^{14}$N) \\ ~[\Kkms] \end{tabular}}
           & \begin{tabular}{c}      $I$(HC$^{15}$N) \\ ~[\Kkms] \end{tabular}
           & $\frac{\hbox{$I$(HC$^{14}$N)}}{\hbox{$I$(HC$^{15}$N)}}$ \\
   \hline
   \uspace1.5 \dspace
   M\,82   & HCN(1--0)   &  88631.602     & 86054.961      &  39.6 & 1.0
                                          & $<$ 0.9 \see{e}&  $>$ 43 \see{e} \\
   \hline
   \end{tabular}
   \end{flushleft}
   {\footnotesize \begin{enumerate} \renewcommand{\labelenumi}{\alph{enumi})}
                                                  \renewcommand{\itemsep}{0pt}
   \item see Skatrud \etal\ (1983)
   \item see Bogey \etal\ (1984)
   \item HNC data from H\"uttemeister \etal\ (1995)
   \item HCN data from Nguyen-Q-Rieu \etal\ (1992)
   \item Limits are 3\,$\sigma$ values.
   \end{enumerate} }
\ \ \\
\end{table*}

\section{Discussion}
\label{sec:CN-Discussion}

   The most stringent limits on \ISOC\ isotope ratios can be deduced
   from CN (see Tables \ref{tbl:CN-line} and \ref{tbl:CN-ISO}).
   So far, little is known on CN emission from the central region of the
   Milky Way, which may serve as a model of what to expect
   in nuclear regions of other galaxies.
   In the following we thus analyse the Galactic center CN data
   (Sect.\,\ref{sec:CN-hyperfine}); in Sect.\,\ref{sec:CN-extragalactic}
   we will then combine these results with those from
   other molecular species to derive isotope ratios in M\,82 and IC\,342.

\subsection{LTE deviations in $^{12}${\rm CN} and $^{13}${\rm CN}}
\label{sec:CN-hyperfine}

   Each \twCN\ rotational energy level with $N > 0$ is split into
   a doublet by spin-rotation interaction.
   Because of the spin of the nitrogen nucleus ($I_1 = 1$),
   each of these components is further split into a triplet of states.
   The \thCN\ energy level scheme is further complicated by
   the spin of the \C13 \ ($I_2 = 1/2$) nucleus.
   The calculated frequencies and relative intensities of the
   resulting hyperfine components are given by
   Skatrud \etal\ (1983) and Bogey \etal\ (1984).

   In order to derive CN optical depths and column densities,
   we can compare the line intensity ratios of various components
   observed with those expected in the optically thin case under
   conditions of Local Thermodynamic Equilibrium (LTE).
   If LTE holds, line saturation leads via
\begin{equation}
   \frac{T_{\rm MB, 1}}{T_{\rm MB, 2}} =
        \frac{1 - {\rm e}^{-\tau_1}}{1 - {\rm e}^{-\tau_2}}\
 \label{eqn:CN-lte}
\end{equation}
   to line intensity ratios which are intermediate between
   those in the optically thin case and unity.
   $T_{\rm MB,i}$ is the observed main beam brightness temperature
   and $\tau_{\rm i}$ is the optical depth of hyperfine component $i$.
   Previous \twCN\ $N$=1--0 observations of dark clouds and star-forming
   regions of the Galactic disk (\eg\ Turner \& Gammon 1975;
   Churchwell 1980; Churchwell \& Bieging 1982, 1983; Crutcher \etal\ 1984)
   did not provide strong evidence for non-LTE effects.

\begin{figure}
   \vspace*{-42 mm}
   \hspace*{-1 cm} \psfig{figure=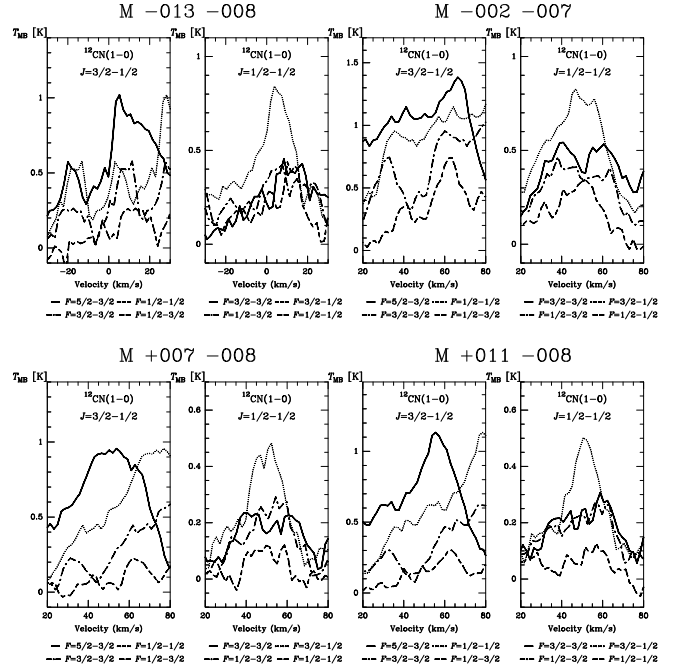,width=10.5 cm}\\
   \vspace*{-1 cm}
   \caption[]
           {Observed spectra of CN $N$=1--0 $J$=3/2--1/2 (left panel) and
            $J$=1/2--1/2 (right panel) toward the indicated clouds
            in the Galactic center region.
            All hyperfine components were obtained simultaneously
            in the same AOS backend; these profiles have been split
            and shifted by their frequency difference.}
 \label{fig:CN-absorb}
\end{figure}

   For optically thin emission under conditions of LTE,
\begin{eqnarray}
   R_{12}\,(^{12}{\rm CN}\ N=1-0)
      & = & \frac{I(J=1/2-1/2)}{I(J=3/2-1/2)} \nonumber \\
      & = & 0.50,
 \label{eqn:CN-r12}
\end{eqnarray}
   where $I$ is the intensity of a given line.
   Our observational data from the Galactic center region
   (Table\,\ref{tbl:CN-depth}), with $R_{12}$ being typically slightly
   smaller than unity but larger than 0.5,
   may provide evidence for \twCN\ line saturation.
   Toward M$-$013$-$008, however, a line intensity ratio
   of $R_{12}$ $\sim$ 1.2 hints at deviations from LTE.
   A detailed examination of our Galactic center data
   (see Table\,\ref{tbl:CN-line} and Fig.\,\ref{fig:CN-absorb}) provides
   further evidence for non-LTE effects:
   The $J$=1/2--1/2 $F$=3/2--3/2 feature that should be
   strongest among the $J$=1/2--1/2 spin-doublet transitions
   is found to be weaker than the $F$=3/2--1/2 component.
   The $J$=1/2--1/2 $F$=3/2--1/2 line is always stronger than the corresponding
   $F$=1/2--3/2 transition even though both lines should have equal strength.
   Another non-LTE effect can be identified when comparing the integrated
   intensity of the weakest resolved \twCN\ hyperfine component, the
   $J$=1/2--1/2 $F$=1/2--1/2 feature, with that of the measured \thCN\ profile.
   With the \twCN\ feature only comprising 1.23\% of the total
   CN $N$=1--0 intensity under optically thin LTE conditions,
   $I$($J$,$F$=1/2,1/2--1/2,1/2)/$I$(\thCN) line intensity ratios
   of 0.35 -- 1.5 would yield \twCN/\thCN\ abundance ratios that are
   well in excess of the Galactic center \ISOC\ value.

   We conclude that lines with $N$=0, $J$,$F$=1/2,1/2 as lower state are
   enhanced relative to lines with $N$=0, $J$,$F$=1/2,3/2 as lower state.
   Apparently, $N$=1--0 hyperfine intensities sensitively
   depend on the relative populations in the $N$=0 doublet.
   A simulation infers that if the $N$=0, $J,F$=1/2,3/2 population
   is `overabundant', all $N$=1--0 features with $N$=0, $F,J$=1/2,1/2
   as lower state show enhanced line temperatures.

   The low LTE intensity of the $J,F$=1/2,1/2--1/2,1/2 transition (1.23\% of
   the total CN $N$=1--0 line strength), the discussion about fractionation
   given below, and the possibility to relate various LTE deviations to a
   single cause favor an explanation based on the relative populations
   of the $N$=0 states over alternative views, either involving
   optically thick \twCN, $N$=0, $J,F$=1/2,1/2--1/2,1/2
   lines or a high degree of isotope selective fractionation.

\begin{table}
   \caption[]
           {Line intensity ratios of \twCN\ and \thCN\ hyperfine components
            toward observed sources}
 \label{tbl:CN-depth}
   \begin{flushleft}
   \begin{tabular}{l r@{$\pm$}l r@{$\pm$}l}
   \hline
\uspace1 \dspace
 Object    & \multicolumn{2}{c}{$R_{12}$ \see{a}}  % &  $\tau_{12}$
           & \multicolumn{2}{c}{$R_{13}$ \see{b}} \\
   \hline
\uspace1
 M\,82         &  0.43 & 0.12  & \multicolumn{2}{c}{\ldots} \\
 IC\,342       &  0.69 & 0.21  & \multicolumn{2}{c}{\ldots} \\
\uspace0.5
 M$-$013$-$008 &  1.18 & 0.05  &  0.50 & 0.07 \\
 M$-$002$-$007 &  0.81 & 0.01  &  0.87 & 0.07 \\
 M+007$-$008   &  0.64 & 0.02  &  0.85 & 0.10 \\
\dspace
 M+011$-$008   &  0.86 & 0.03  &  0.77 & 0.11 \\
%              &    2&75  & 0.19 &   3&56  & 0.24 &  0.77 & 0.11 \\
   \hline
   \end{tabular}
   \end{flushleft}
   {\footnotesize \begin{enumerate} \renewcommand{\labelenumi}{\alph{enumi})}
                                                  \renewcommand{\itemsep}{0pt}
   \item $R_{12}$ is the integrated intensity ratio of the $J$=1/2--1/2
         to $J$=3/2--1/2 spin-doublet lines.
         In the optically thin case, the LTE value is 0.50
         (see Eq.\,(\ref{eqn:CN-r12})).
   \item $R_{13}$ is the integrated intensity ratio of the $F_1$=1 $F_2$=2--1
         to the $F_1$=1 $F_2$=1--1 and $F_1$=0 $F_2$=1--0 components.
         In the optically thin case under LTE conditions, the value is 0.83
         (see Eq.\,(\ref{eqn:CN-r13})).
   \end{enumerate} }
\end{table}

   For \thCN, the assumption of optically thin emission is very plausible.
   As already mentioned, \thCN\ hyperfine components have been resolved
   into two main groups of transitions (Fig.\,\ref{fig:CN-GC}).
   We can estimate the deviation from (optically thin) LTE intensity ratios
   by examining $R_{13}$, for which we obtain under LTE conditions
\begin{eqnarray}
   R_{13}(^{13}{\rm CN}\ N=1-0)
      & = & \frac{I(F_2=2-1)}{I(F_2=1-1 + F_2=1-0)} \nonumber \\
      & = & 0.83.
 \label{eqn:CN-r13}
\end{eqnarray}
   In Table \ref{tbl:CN-depth} the integrated line intensity ratios
   in the observed Galactic center clouds are listed.
   Most of these ratios are, within the 1$\sigma$ error,
   consistent with the value from Eq.\,(\ref{eqn:CN-r13}).
   The exception is M$-$013$-$008, where the ratio is 0.50 $\pm$ 0.07.

   Integrating over all hyperfine components, the CN data from the Galactic
   center (Table \ref{tbl:CN-line}) show that $I$(\twCN)/$I$(\thCN) = 9 -- 15.
   Based on data from a variety of sources and
   molecular species (\CeiO, \FORM, \HCOp,
   NH$_2$CHO, OCS; see Wannier 1980;
   Wilson \& Rood 1994, and many references therein), there is
   strong evidence for an actual \ISOC\ ratio of 25 $\pm$ 5
   in the interstellar medium of the Galactic center region.
   Accounting for possible line saturation in the \twCN\ profiles,
   line intensity and isotope ratios are therefore consistent.
   Our CN data provide {\it lower limits} to the
   \ISOC\ isotope ratio and neither fractionation nor isotope selective
   photoionization play a dominant role.

   Fractionation can be analysed in some more detail:
   The \twCO /\thCO\ abundance ratio is affected by the reaction
\begin{equation}
   ^{13}{\rm C}^{\rm +} +\ ^{12}{\rm CO}\; \longrightarrow \;
   ^{12}{\rm C}^{\rm +} +\ ^{13}{\rm CO} +\ \Delta E_{\rm 35\,K}
 \label{eqn:CN-reaction}
\end{equation}
   (Watson \etal\ 1976), which enhances \thCO\ relative to \twCO\ and
   $^{12}$C$^{\rm +}$ relative to $^{13}$C$^{\rm +}$ in the more
   diffuse \Cp\ rich parts of molecular clouds.
   For \twCN\ and \thCN, the corresponding difference in the energies
   of the ground vibrational states is $\sim$ 32\,K.
   While reaction (\ref{eqn:CN-reaction}) has little effect on the
   \HtwCOp/\HthCOp\ abundance ratio, other carbon-bearing molecules like
   HCN, HNC, \FORM, and CN are affected in the sense that their \ISOC\
   abundance ratios follow \Cp\ and are thus larger than the corresponding
   isotope ratio (Langer \etal\ 1984; Langer \& Graedel 1989).
   With \CeiO\ and \FORM\ bracketing the actual carbon isotope ratio,
   observations show that these deviations must be small
   ($\sim$ 20 -- 30\% in the Galactic disk;
   cf.\ Henkel \etal\ 1982; Langer \& Penzias 1990).
   In Galactic center molecular clouds \Tkin\ is
   larger than in clouds of the disk.
   \Tkin\ may be largest in the more active regions
   of IR-luminous galaxies (\eg\ Mauersberger \etal\ 1986; Ho \etal\
   1990; Solomon \etal\ 1992) and isotope selective CN
   fractionation should thus be negligible.

\subsection{Extragalactic isotope ratios}
\label{sec:CN-extragalactic}

\subsubsection{$^{12}$C/$^{13}$C ratios}
\label{sec:CN-isoc}

   Our extragalactic CN data can be used to estimate \ISOC\ isotope ratios,
   (1) because the spectra from the Galactic center region provide
       `reasonable' results (see Sect.\,\ref{sec:CN-hyperfine}) and
   (2) because significant non-LTE effects are not
       seen in the M\,82 and IC\,342 spectra.
   The $R_{12}$ line intensity ratios are close to the LTE value (see
   Eq.\,(\ref{eqn:CN-r12}) and Table \ref{tbl:CN-depth}).
   Furthermore, the HC$^{13}$CCN and HCC$^{13}$CN features, observed in the
   Galactic center \thCN\ spectra (Fig.\,\ref{fig:CN-GC}), are not expected
   to disturb our extragalactic \thCN\ profiles: HC$_3$N is extremely weak
   in extragalactic sources (\eg\ Henkel \etal\ 1988; Mauersberger \etal\ 1990).
   Toward our extragalactic sources, \twCN\ is thus
   likely optically thin and the $I$(\twCN)/$I$(\thCN) line intensity ratios of
   $>$ 40 and $>$ 30 for M\,82 and IC\,342, respectively, can be interpreted
   as lower limits to the \ISOC\ isotope ratio.

   In Table \ref{tbl:CN-ISO}, lower limits to the \ISOC\ intensity ratio are
   displayed for a variety of molecular species toward M\,82 and IC\,342.
   Unlike the $N$=1--0 CN transitions, other observed molecular lines
   are not significantly broadened due to hyperfine splitting.
   The emission from the \C13 -bearing isotopic species is probably optically
   thin, while the emission from the main species may be saturated.
   Since the line intensity ratio of the common \C12 \ to its rare
   \C13 -bearing isotopic species will be reduced by optical depth and
   self-absorption effects (this is not fully compensated by differences in
   excitation; see Henkel \etal\ 1994), the lower limit of the intensity ratio
   is also the lower limit to the \ISOC\ isotopic abundance ratio.

   Combining our result with previous work done toward NGC\,253 and NGC\,4945
   (Henkel \etal\ 1993, 1994), a small \ISOC\ ratio as observed in our
   Galactic center region appears not to be typical for galaxies with
   infrared luminosities similar to or higher than that of
   the nuclear region of the Milky Way.
   This result suggests (Casoli \etal\ 1992b) that
   (1) gas inflow from the outer regions adds large amounts of low-metallicity,
       \C13 -poor material to the nuclear interstellar medium or that
   (2) \C12 \ may be more efficiently produced (relative to \C13 ) in regions
       with a high rate of massive star formation.

\subsubsection{Limits to the \ISOA\ ratio}
\label{sec:CN-isoa}

   OH and \FORM\ data indicate a \ISOA\ ratio of 250 $\pm$ 30
   for the Galactic center region (\eg\ Wilson \& Rood 1994).
   Combining \twCO, \thCO, and \CeiO\ data with our upper limits
   to the \ISOC\ ratio deduced in Sect.\,\ref{sec:CN-isoc},
   we can also constrain the \ISOA\ ratio.
   Following Henkel \etal\ (1994; their Sect.\,6.2) and combining
   IRAM 30-m \twCO, \thCO, and \CeiO\ data from Eckart \etal\ (1990),
   Loiseau \etal\ (1990), Sage \etal\ (1991), Wild \etal\ (1992),
   and G\"usten \etal\ (1993), we find for M\,82
\begin{equation}
   ^{12}{\rm C}/^{13}{\rm C} \leq  [I(^{12}{\rm CO})/I(^{13}{\rm CO}]
   \tau (^{12}{\rm CO}) \sim 17\,\tau (^{12}{\rm CO})
\end{equation}
   and
\begin{equation}
   ^{16}{\rm O}/^{18}{\rm O} \leq  [I(^{12}{\rm CO})/I({\rm C}^{18}{\rm O}]
   \tau (^{12}{\rm CO}) \sim 44\,\tau (^{12}{\rm CO})
\end{equation}
   and for IC\,342
\begin{equation}
   ^{12}{\rm C}/^{13}{\rm C} \leq  [I(^{12}{\rm CO})/I(^{13}{\rm CO}]
   \tau (^{12}{\rm CO}) \sim 11\,\tau (^{12}{\rm CO})
\end{equation}
   and
\begin{equation}
   ^{16}{\rm O}/^{18}{\rm O} \leq  [I(^{12}{\rm CO})/I({\rm C}^{18}{\rm O}]
   \tau (^{12}{\rm CO}) \sim 51 \tau (^{12}{\rm CO}).
\end{equation}
   This implies for M\,82 $\ISOC \sim 0.39 \times (\ISOA)$ and
   for IC\,342 $\ISOC \sim 0.22 \times (\ISOA)$,
   while for the Galactic center region, $\ISOC \sim 0.10 \times (\ISOA)$.
   With the lower \ISOC\ limits deduced in Sect.\,\ref{sec:CN-isoc} and
   accounting for uncertainties in the C$^{18}$O line intensities,
   we thus find \ISOA\ $>$ 90 and $>$ 125 for M\,82 and IC\,342, respectively.

\subsubsection{The \ISON\ ratio in M\,82}
\label{sec:CN-ison}

   The \ISON\ abundance ratio (270 in our solar system and $\sim$ 400
   in the local interstellar medium; \eg\ Dahmen \etal\ 1995)
   has so far not been measured in an extragalactic object.
   According to Henkel \& Mauersberger (1993), such ratios may be small
   ($\la$ 150) in nuclear starburst regions.
   Combining \HCfiN\ $J$=1--0 data from M\,82 with HCN (\ie\ \HCfoN),
   we can derive a lower limit to the \ISON\ abundance ratio.
   From Table \ref{tbl:CN-ISO} a lower 3$\sigma$ limit of 43 is
   obtained for the $I$(\HCfoN)/$I$(\HCfiN) ratio.
   Estimating an \HCfoN\ optical depth of $\sim$ 3 -- 4
   (Henkel \etal\ 1993), a lower limit of 43 from the $I$(\HCfoN)/$I$(\HCfiN)
   intensity ratio implies that the \ISON\ isotope ratio should exceed 100.

\section{Conclusions}
\label{sec:CN-Conclusions}

   Having studied the $N$=1--0 line profiles of \twCN\ and \thCN\ toward
   four Galactic center clouds and a variety of molecular species toward the
   central region of M\,82 and IC\,342, we obtain the following main results:

\begin{enumerate} \renewcommand{\labelenumi}{(\arabic{enumi})}
   \item In the Galactic center clouds,
         non-LTE populations in the $^{12}$CN $N$=0 doublet lead to
         enhanced line intensities for hyperfine transitions with $N$=0,
         $J$,$F$=1/2,1/2 as lower state.
         Alternative explanations for the LTE deviations,
         either involving opaque $J,F$=1/2,1/2--1/2,1/2 lines or
         significant isotope selective fractionation are not attractive.
         No signs for line saturation or non-LTE effects are
         found in the CN spectra from M\,82 and IC\,342; non-LTE intensities
         in the \thCN\ $N$=1--0 profile are only seen toward one Galactic
         center cloud, M--013--008.

   \item If line saturation affects the observed \twCN\ features,
         $I$(\twCN)/$I$(\thCN) line intensity ratios should be
         smaller than the actual \ISOC\ abundance ratio, which is $\sim$ 25
         in the Galactic center region.
         With $I$(\twCN)/$I$(\thCN) = 9 -- 15, some saturation should be
         present.
         Since $I$(\twCN)/$I$(\thCN) $\la $ \ISOC, we find no reason to avoid
         CN as a tracer to contrain \ISOC\ isotope ratios in
         the nuclear regions of nearby galaxies.

   \item $^{12}$C/$^{13}$C limits from CN are $>$ 40 and $>$ 30 for M\,82
         and IC\,342, respectively.
         Since the \ISOC\ ratios have been found to be $\sim$ 40, $\sim$ 50,
         and 25 $\pm$ 5 in NGC\,253, NGC\,4945, and the Galactic center
         region, these results are consistent with large \ISOC\ ratios
         ($>$25) in the nuclear regions of starburst galaxies (M\,82,
         NGC\,253, and NGC\,4945).

   \item Combining these results with measurements of various CO isotopic
         species, \ISOA\ $>$ 90 and $>$ 125 in M\,82 and IC\,342, respectively.

   \item A lower \ISON\ limit of 100 is estimated, based on an
         HCN $J$=1--0 optical depth of $\sim$ 3 -- 4 in the main species.
\end{enumerate}

\begin{acknowledgements}
   We thank M.~Gu\'elin for critically reading the manuscript.
   Y.-N.~Chin thanks for financial support through DAAD
   (Deutscher Akademischer Austauschdienst) grant \hbox{573~307~0023} and
   through National Science Council of Taiwan grant \hbox{86-2112-M001-032}.
   R.~Mauersberger acknowledges support as a Heisenberg fellow by
   the Deutsche Forschungsgemeinschaft (DFG).
\end{acknowledgements}


\begin{thebibliography}{}
 \bibitem{xyz}
   Aalto S., Black J.H., Johansson L.E.B., Booth R.S., 1991, A\&A 249, 323
 \bibitem{xyz}
   Baars J.W.M., Hooghoudt B.G., Mezger P.G., de Jonge M.J., 1987, A\&A 175, 319
 \bibitem{xyz}
   Bogey M., Demuynck C., Destombes J.L., 1984, Can. J. Phys. 62, 1248
 \bibitem{xyz}
   Booth R.S., Delgado G., Hagstr\"om M., \etal, 1989, A\&A 216, 315
 \bibitem{xyz}
   Casoli F., Dupraz C., Combes F., 1992a, A\&A 264, 49
 \bibitem{xyz}
   Casoli F., Dupraz C., Combes F., 1992b, A\&A 264, 55
 \bibitem{xyz}
   Churchwell E., 1980, ApJ 240, 811
 \bibitem{xyz}
   Churchwell E., Bieging J.H., 1982, ApJ 258, 515
 \bibitem{xyz}
   Churchwell E., Bieging J.H., 1983, ApJ 265, 216
 \bibitem{xyz}
   Combes F., Casoli F., Encrenaz P., Gerin M., Laurant C., 1991, A\&A 248, 607
 \bibitem{xyz}
   Crutcher R.M., Churchwell E., Ziurys L.M., 1984, ApJ 283, 668
 \bibitem{xyz}
   Dahmen G., Wilson T.L., Matteucci F., 1995, A\&A 295, 194
 \bibitem{xyz}
   Downes D., 1989, Introductory Course in Galaxies' Evolution and
     Observational Astronomy, eds. I. Appenzeller, H. Habing, P. Lena,
     Springer, Heidelberg, p353
 \bibitem{xyz}
   Eckart A., Downes D., Genzel R., \etal, 1990, ApJ 348, 434
 \bibitem{xyz}
   G\"usten R., Serabyn E., Kasemann C., \etal, 1993, ApJ 402, 537
 \bibitem{xyz}
   Henkel C., Mauersberger R., 1993, A\&A 274, 730
 \bibitem{xyz}
   Henkel C., Wilson T.L., Bieging J.H., 1982, A\&A 109, 344
 \bibitem{xyz}
   Henkel C., Wouterloot J.G.A., Bally J., 1986, A\&A 155, 193
 \bibitem{xyz}
   Henkel C., Mauersberger R., Schilke, P., 1988, A\&A 201, L23
 \bibitem{xyz}
   Henkel C., Baan W.A., Mauerberger R., 1991, A\&AR 3, 47
 \bibitem{xyz}
   Henkel C., Mauersberger R., Wiklind T., \etal, 1993, A\&A 268, L17
 \bibitem{xyz}
   Henkel C., Whiteoak J.B., Mauersberger R., 1994, A\&A 284, 17
 \bibitem{xyz}
   Ho P.T.P., Martin R.N., Turner J.L., Jackson J.M., 1990, ApJ
   355, L19
 \bibitem{xyz}
   H\"uttemeister S., Henkel C., Mauersberger R., \etal, 1995, A\&A 295, 571
 \bibitem{xyz}
   Loiseau N., Nakai N., Sofue Y., \etal, 1990, A\&A 228, 331
 \bibitem{xyz}
   Langer W.D., Graedel T.E., 1989, ApJS 69, 241
 \bibitem{xyz}
   Langer W.D., Graedel T.E., Frerking M.A., Armentrout P.B., 1984, ApJ 277, 581
 \bibitem{xyz}
   Langer W.D., Penzias A.A., 1990, ApJ 357, 477
 \bibitem{xyz}
   Mauersberger R., Henkel C., Wilson T.L., Walmsley C.M., 1986, A\&A 162, 199
 \bibitem{xyz}
   Mauersberger R., Henkel C., Sage, L.J., 1990, A\&A 236, 63
 \bibitem{xyz}
   Mauersberger R., Gu\'elin M., Mart\'{\i}n-Pintado J., \etal,
     1989, A\&AS 79, 217
 \bibitem{xyz}
   Nguyen-Q-Rieu, Jackson J.M., Henkel C., Truong-Bach, Mauersberger R.,
     1992, ApJ 399, 521
 \bibitem{xyz}
   Renzini A., Voli M., 1981, A\&A 94, 175
 \bibitem{xyz}
   Sage L.J., Mauersberger R., Henkel C., 1991, A\&A 249, 31
 \bibitem{xyz}
   Skatrud D.D., de Lucia F.C., Blake G.A., Sastry K.V.L.N.,
     1983, \JMolSp\ 99, 35
 \bibitem{xyz}
   Solomon P.M., Downes D., Radford S.J.E., 1992, ApJ 398, L29
 \bibitem{xyz}
   Turner B.E., Gammon R.H., 1975, ApJ 198, 71
 \bibitem{xyz}
   Wannier P.G., 1980, ARA\&A 18, 399
 \bibitem{xyz}
   Watson W.D., Anichich V.G., Huntress W.T., 1976, ApJ 205, L165
 \bibitem{xyz}
   Wild W., Harris A.I., Eckart A., \etal, 1992, A\&A 265, 447
 \bibitem{xyz}
   Wilson T.L., Rood R.T., 1994, ARA\&A 32, 191
\end{thebibliography}
\end{document}